\documentstyle[12pt]{article}
\begin{document}
\input epsf.tex
\def\a{\alpha}
\def\b{\beta}
\def\ch{\chi}
\def\d{\delta}
\def\e{\epsilon}
\def\E{{\cal E}}
\def\f{\phi}
\def\g{\gamma}
\def\h{\eta}
\def\i{\iota}
\def\j{\psi}
\def\k{\kappa}
\def\l{\lambda}
\def\m{\mu}
\def\n{\nu}
\def\o{\omega}
\def\p{\pi}
\def\q{\theta}
\def\r{\rho}
\def\s{\sigma}
\def\t{\tau}
\def\u{\upsilon}
\def\x{\xi}
\def\z{\zeta}
\def\D{\Delta}
\def\F{\Phi}
\def\G{\Gamma}
\def\J{\Psi}
\def\L{\Lambda}
\def\O{\Omega}
\def\P{\Pi}
\def\S{\Sigma}
\def\U{\Upsilon}
\def\X{\Xi}
\def\T{\Theta}
\def\vf{\varphi}
\def\ve{\varepsilon}
\def\cC{{\cal X}}
\def\cD{{\cal Y}}
\def\Ab{\bar{A}}
\def\gi{g^{-1}}
\def\li{{ 1 \over \l } }
\def\lb{\l^{*}}
\def\zb{\bar{z}}
\def\ub{u^{*}}
\def\vb{v^{*}}
\def\Tb{\bar{T}}
\def\sech {{\rm{sech}}}
\def\csch {{\rm{csch}}}
\def\cn{{\rm{cn}}}
\def\dn{{\rm{dn}}}
\def\sn{{\rm{sn}}}
\def\be{\begin{equation}}
\def\ee{\end{equation}}
\def\ben{\begin{eqnarray}}
\def\een{\end{eqnarray}}
\def\lt{\tilde{\lambda}}
\def\ben{\begin{eqnarray}}
\def\een{\end{eqnarray}}
\def\pb{\bar \partial}
\def\pp{\partial}
\def\s{\sigma}
\def\cw{\psi^{cw}}
\def\Et{\tilde{E}}
\def\Im{{\rm{Im}}}
\def\Re{{\rm{Re}}}
\hsize=15truecm
\addtolength{\topmargin}{-0.6in}
\vsize=26.5truecm
\hoffset=-0.3in

\rightline{\today} 
\vskip2cm
\centerline{\Large\bf Sine-Gordon Soliton on a Cnoidal Wave Background }
\vskip 1cm
\centerline{
H. J. Shin\footnote{ Electronic address; hjshin@khu.ac.kr }}
\vskip 3mm
\centerline{Department of Physics and Research Institute of Basic
Science}
\centerline{Kyung Hee University, Seoul 130-701,  Korea}
\vskip 2cm
\centerline{\bf ABSTRACT}
\vskip 5mm

The method of Darboux transformation, which is applied on cnoidal wave solutions
of the sine-Gordon equation, gives solitons moving on a cnoidal wave background.
Interesting characteristics of the solution, i.e.,  the velocity of solitons and  
the shift of crests of cnoidal waves along a soliton,
are calculated.
Solutions are classified into three types
(Type-1A, Type-1B, Type-2) according to their apparent distinct properties.

\vskip 5mm

\newpage

\setcounter{footnote}{0}

\vskip 5mm

\section{Introduction}
The sine-Gordon equation is important for a number of physical and mathematical systems where topological
solitons are present. In the case of  physical applications, it is a canonical model
describing charge density waves \cite{density}, fluxon dynamics in Josephson junctions \cite{fluxon},
and DNA promoter dynamics \cite{dna}, for example.
In the case of mathematical topics, we can mention the embedding of a surface with constant
negative curvature in the 3-dimensional space \cite{embed}.
In theoretical physics, it was studied as a model system of hadrons with the idea that
extended particles in quantum field theory can be associated with soliton solutions
of integrable equations \cite{das}. The quantum sine-Gordon model is also developed and
the exact S-matrix of the solitons was derived by using the bootstrap methods \cite{zam}.

The important property of the sine-Gordon equation in these studies is the existence of soliton solutions.
The simple functional form of N-soliton has led to a wide range of scientific applications.
Considering the potential applicability of the soliton solutions, it is desired to have a more generalized form of 
soliton solutions, having more parameters and characteristics. One possible scheme in this direction
is constructing soliton solutions lying on a background field.
It would give more freedom in controlling and characterizing solitons
by using proper backgrounds.
Indeed, the sine-Gordon equation, as an integrable equation, has solutions of superposed states ``soliton + cnoidal wave".
Recently, this type of solutions have attracted new interest in nonlinear optics, described by the nonlinear Schr\"{o}dinger
equation (NLSE), due to the possibility of dynamically reconfigurable photonic structures \cite{des,kivshar,jph}.

In fact, the ``soliton + cnoidal wave" solution was obtained from the general quasi-periodic solutions
of N-phase theta functions \cite{koz,date},
by taking degenerate limit of the 2-phase solution \cite{jaw1,jaw2}. 
To apply solutions from the N-phase theta functions
to real physical problems, the authors in \cite{jaw1,jaw2} solve
the so-called ``effectivization" problem, which is related to extracting physical solutions by
taking proper initial conditions \cite{jaw1,jaw2,kam,shin11}. However we find that above solutions, though explicit, are
inconvenient to analyze their properties, as the parameters in the solutions are related in a complicated form.
In fact, the mathematical and computational complexities of computing N-phase theta functions
have hindered the use of these solutions in applications except some clever calculations like in \cite{novi,sin}.

In this paper, we employ a simple, but powerful soliton
finding technique \cite{kivshar,jph} based on the Darboux transformation (DT) to find solutions of ``soliton + cnoidal wave".
Compared to the results from the N-phase theta functions, the form of solution, especially the parameters, are simple
and convenient to analyze various properties of the solutions. Explicit real
solutions are described by a parameter $u$ without any further constraints and it
is directly related to the characteristics of solutions. This DT technique gives a clear classification scheme of the solutions
(Type-1A, Type-1B, Type-2) according to their apparent distinct properties like the modulating form of cnoidal
waves and the velocities of solitons. The velocity of solitons are expressed in a form of relativistic addition, 
which is a unique feature of the
``soliton + cnoidal wave" solution of the {\it relativistic} sine-Gordon equation and is essentially different
from that of the {\it nonrelativistic} NLSE and KdV equation.
As the sine-Gordon theory has various physical applications, our solution could yield a new interest
on controlling the soliton using background fields.

\section{sine-Gordon equation}
\subsection{Cnoidal wave solutions}
The sine-Gordon equation is
\be
\pp \pb \phi = 2 \b \sin 2 \phi,
\label{dnls}
\ee
where $\pp = {\pp \over \pp z}, \pb = {\pp \over \pp \zb}$, and
$\zb \equiv t-x$ and $z \equiv t+x$ represent the lightcone coordinates,
while $x, t$ are ordinary space-time coordinates.
It has two types of cnoidal wave solutions;
\ben
\f_c^{(1)}(z, \zb)=\sin^{-1} ({\rm sn} (\ch,k)), ~\pp \f_c^{(1)}={2 \over k} \sqrt{\b \over V} \dn(\ch,k) 
,~&&\rm{Type-1} \nonumber \\
\f_c^{(2)}(z, \zb)=\sin^{-1} (k ~{\rm sn} (k \ch,k)), ~\pp \f_c^{(2)}=2 k \sqrt{\b \over V} \cn(k \ch,k), 
~&&\rm{Type-2}
\label{cnoid}
\een
where
\be
\ch= {2 \over k} \sqrt{\b \over V} (z-V \zb) ={4 \over k} \sqrt \b {x-v_c t \over \sqrt{1-v_c ^2}},
\label{vel}
\ee
and sn, dn are the standard Jacobi elliptic functions.
$V$ is related to the velocity $v_c$ of
the cnoidal wave as $V=(1+v_c) /(1-v_c)$ and $k \in (0,1)$ is the modulus of the Jacobi function.
Here we consider the case of $V>0, |v_c|<1$ and $\b>0$. The $V<0, |v_c|>1$ case, corresponding to
the rescaled inverted pendulum \cite{sin}, can be obtained by considering the
$x \leftrightarrow t, \b \leftrightarrow -\b$ symmetry of the sine-Gordon theory.
As far as elliptic functions are involved we employ terminology and
notation of Ref. \cite{jms} without further explanations.
$\f_c^{(1)}$ of the Type-1 solution is a monotonically increasing or decreasing function of $\ch$, while
$\f_c^{(2)}$ of the Type-2 is an oscillating function of $\ch$.

\subsection{Lax pair}
To obtain a superposed ``soliton +  cnoidal wave"  solution using the DT method \cite{dt,Park1,shin4}, we need to find
a solution of the following linear equations associated to the sine-Gordon equation (Lax pair),
\ben
(\pp +i \b \l) s_1 +i (\pp \f_c^{(i)}) s_2 = 0,~~
(\pp -i \b \l) s_2 +i (\pp \f_c^{(i)}) s_1 = 0, \nonumber \\
(\pb - {i \over \l} \cos (2 \f_c^{(i)}) ) s_1 +{1 \over \l} \sin (2 \f_c^{(i)}) s_2 = 0, \nonumber \\
(\pb +{i \over \l} \cos (2 \f_c^{(i)}) ) s_2 -{1 \over \l} \sin (2 \f_c^{(i)}) s_1 = 0,
\label{linear}
\een
where $\l $ is an pure imaginary number and $i=1,2$ represents two types of cnoidal solutions  in Eq. (\ref{cnoid}).
We note that the compatibility between  the equations (\ref{linear}) requires $\f_c^{(i)}$ satisfies the Eq. (\ref{dnls}).

\subsection{Sym's solution}
The  solution of the linear equations (\ref{linear}) for the cnoidal wave ($\f_c^{(i)}, i=1,2$) can be obtained 
by a slight modification of Sym's solution \cite{sym}.  Sym first introduced this type of solution in a context
of vortex motion in hydrodynamics.
It was then applied to NLSE-related problems in Ref. \cite{shin2,kivshar,jph}. A more detailed proof of Sym's solution
(in a slightly different notation) is given
in the Appendix of Ref. \cite{jph}. The Sym's solution in the case of sine-Gordon equation
for the Type-1 cnoidal solution, $\f_c^{(1)}$, is (to be called as the Type-1A solution)
\ben
s_1 &=& \{a e^M \Theta_s(\ch -u) + ib{\sn(u,k) \over \cn(u,k) \dn(\ch +u,k)} e^{-M} \Theta_s(\ch +u) \} /\Theta_s(\ch),
\nonumber \\
s_2 &=& \{ - ia {\sn(u,k) \over \cn(u,k) \dn(\ch -u,k)} e^M \Theta_s(\ch -u) +b e^{-M} \Theta_s(\ch +u) \}  /\Theta_s(\ch),
\label{s12}
\een
where $a, b$ are arbitrary real parameters.
Here,
\be
M = \sqrt{V \b} {\cn (u,k)^4 +(k^2 -1) \sn (u,k)^4 \over k \sn (u,k) ~\cn (u,k)  \dn (u,k) } \zb +
 \left( {\Theta_s '(u) \over \Theta_s(u)} +{\dn (u,k) (1 - 2 \sn (u,k)^2 ) \over 2 \sn (u,k) ~\cn (u,k)} \right) \ch,
\label{MN}
\ee
and
\be
\Theta_s (u) =\theta_0 ( {\pi u \over 2 K}) = 1 + 2 \sum (-)^n q^{n^2} \cos ({n \pi u \over K}),
\ee
with $q =\exp (-\p K' / K)$ and Jacobi theta function $\theta_0$.
The parameter $u$ is related to $\l$ as
\be
\l ={i \over \sqrt{V \b}} {\dn (u,k) \over k  ~\cn (u,k) ~\sn (u,k)  }.
\label{ol}
\ee

\section{Type-1A soliton solution on a dn-type background}
\subsection{soliton solution}
A new solution $\f_{c-s}$, describing a soliton moving on a cnoidal wave $\f_c^{(1)}$, can
be constructed using the DT method \cite{dt,Park1,shin4} as following;
\ben
\pp \f_{c-s} (z,\zb) = \pp \f_c^{(1)} (z,\zb) -4\b \l {s_1 s_2 \over s_1 ^2-s_2 ^2}, \nonumber \\
\sin (2 \f_{c-s})= 4i {s_1 s_2 (s_1 ^2+s_2 ^2) \over (s_1 ^2-s_2 ^2)^2 } \cos (2 \f_c^{(1)}) 
-{s_1 ^4+6 s_1 ^2 s_2 ^2+s_2 ^2 \over (s_1 ^2-s_2 ^2)^2}  \sin (2 \f_c^{(1)}).
\label{newj}
\een
Using the fact that $s_i, i=1,2$ in Eq. (\ref{s12}) satisfy the associated linear equations (\ref{linear}), it can be proved
that $\f_{c-s}$ in Eq. (\ref{newj}) satisfies the sine-Gordon equation (\ref{dnls}).
It can be seen that the reality of $\f_{c-s}$ requires $u$, as well as $a, b$, are real in the Type-1A solution.

Figure 1(a) plots $\pp \f_{c-s}$ as a function of time $t$ and space $x$. It is obtained by using 
Eqs. (\ref{newj}), (\ref{cnoid}), (\ref{s12}) and (\ref{MN}).
It shows a characteristic 
soliton of sine-Gordon equation lying on a cnoidal wave background.
The parameters used are $V=1, k=0.8, u=0.7, \b=1, a=b=1$.
Figure 1(b) shows a time sliced view of the soliton in Figure 1(a) at $t=0$.
These figures are drawn using MATHEMATICA, which is also used to check that
the solution in Eq. (\ref{newj}) indeed satisfies the equation of motion (\ref{dnls}).
\begin{figure}
\leftline{\epsfxsize 2.6 truein \epsfbox {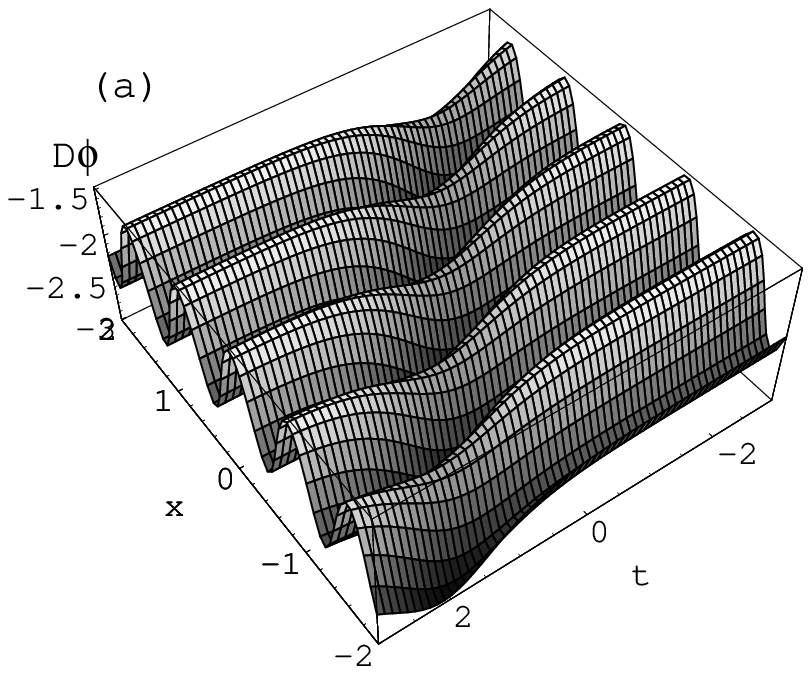}}
\vglue -1.9 in
\rightline{\epsfxsize 2.6 truein \epsfbox {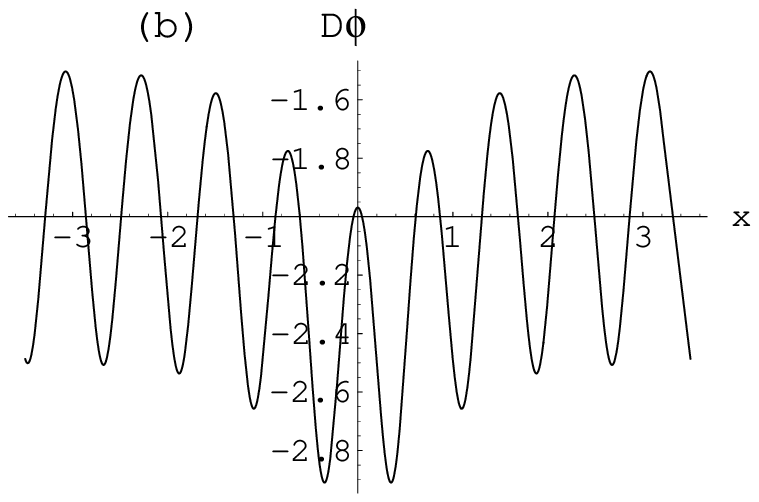}}
\caption{$\pp \f_{c-s}$ showing a sine-Gordon soliton residing on a cnoidal wave background (Type-1A solution);
(a) 3-dimensional plot, (b) time-sliced view at $t=0$. The parameters are $V=1, k=0.8, u=0.7, \b=1, a=b=1$.}
\end{figure}

\subsection{Properties of the solution}
\subsubsection{Shift of cnoidal crests}
In figure 1, we can see that $\f_{c-s}$ becomes 
a cnoidal wave when we move away from the soliton. In fact, as $M \rightarrow \infty$,
$s_2 \rightarrow -i \sn(u,k)  \dn(\ch -u,k) s_1/ \cn(u,k) $ and
$\pp \f_{c-s} \rightarrow -2 \sqrt{\b / V} \dn(\ch -2u,k)/k$, which is a cnoidal wave.
Similarly, in the region $M \rightarrow -\infty$,
$s_2 \rightarrow -i \cn(u,k) s_1/ (\dn(\ch +u,k) \sn(u,k)) $ and
$\pp \f_{c-s} \rightarrow -2 \sqrt{\b / V} \dn(\ch +2u,k)/k$.
This calculation shows that crests of the
cnoidal wave are shifted by $4u$ across the soliton, which can be seen in Fig. 1.
Explicitly, the shift of crests for parameters of Fig. 1 is $4 u=2.8$ in $\ch$ and 
$k u=0.56$ ($V=1, v_0=0, \b=1$) in $x$, use Eq. (\ref{vel}). 

\subsubsection{velocity of the soliton}
The soliton in Fig. 1 moves along a line described by $M=0$, which can be casted in a form $0=z+\D \zb$ with
\be
\D=-V +V {\cn (u,k)^4 +(k^2 -1) \sn (u,k)^4 \over \sn (u,k) ~\cn (u,k)  \dn (u,k) }
/\left( 2{\Theta_s '(u) \over \Theta_s(u)} +{\dn (u,k) (1 - 2 \sn (u,k)^2 ) \over \sn (u,k) ~\cn (u,k)}  \right).
\ee
Using the space-time coordinates $t, x$ instead of the lightcone coordinates $z, \zb$, it is written as $0=z+\D \zb=2 (x -v t)/(1-v)$
where $v= (\D+1)/(\D-1)$. It shows $v$ is the velocity of the soliton in the ordinary space-time coordinates.
We can express the soliton velocity $v$ in terms of the $v_c$ ($V=(1+v_c) /(1-v_c)$) of the cnoidal wave as $v=(v_c+v_0)/(1+v_c v_0)$ where
\be
v_0 =v (v_c=0) = {k^2 \sn^2 u ~\cn^2 u +\dn^2 u~(1-2 \sn^2 u) \over -4 \sn u~\cn u~\dn u~ \Theta_s '(u) / \Theta_s(u)
+k^2 \sn^2 u~\cn^2 u-\dn^2 u ~(1-2 \sn^2 u)},
\label{vs0}
\ee
where $\sn u = \sn(u,k)$, etc.
It shows that $v_0$ is the intrinsic velocity of the soliton observed in the moving frame of the cnoidal wave,
which depends on the parameters $u$ and $k$.
The soliton velocity $v$ is given by the relativistic addition of two velocities, 
$v_c$ (velocity of the cnoidal wave) and $v_0$ (intrinsic velocity of the soliton), where $v_c$ does not depend on $k$ or $u$.
Figure 2 shows the intrinsic velocity $v_0$ in $u$ for a soliton on cnoidal wave backgrounds of $k=0.5$ (solid line), and $k=0.9$ (dotted line).
\begin{figure}
\centerline{\epsfxsize 3. truein \epsfbox {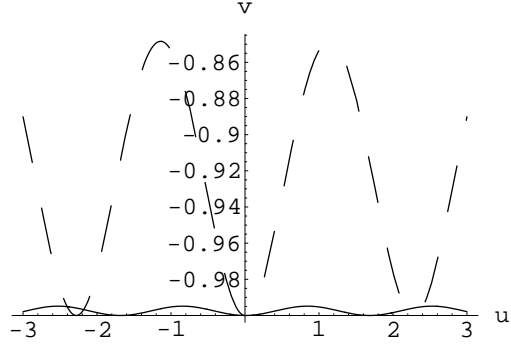}}
\caption{Intrinsic velocity $v_0$ versus $u$, Type-1A sine-Gordon soliton for $k=0.5$ (solid line) and $k=0.9$ (dashed line).
}
\end{figure}
Note that $v_0$ is a periodic function in $u$ with the periodicity $K$. ($K(k=0.5)=1.69, K(k=0.9)=2.28$)
It shows that the intrinsic velocity $v_0 = -1$ at $u=n K$ for integer $n$.
Using Eq. (\ref{vs0}) and the relation
\be
E(\sin^{-1} (\sn u)) ={\Theta_s '(u) \over \Theta_s(u)}+{E \over K} u,
\ee
we can find that the velocity $v_0 = -k' K/E$ at $u=K/2$. Especially, this value is
$-0.995$ for $k=0.5$, $-0.848$ for $k=0.9$, $-1$ for $k=0$, and $0$ for $k=1$. Thus a soliton can be stopped only
when it resides on a cnoidal wave with $k=1$.
On a $k=0$ background, the soliton moves with a velocity $v_0 =-1$.
Note that there is a symmetry, $t \rightarrow -t$, in the ``soliton + cnoidal wave" solution in Eq. (\ref{newj}).
This symmetry gives a solution having a soliton velocity $-v_0$.

\section{Type-1B soliton solution on a dn-type background}
\subsection{soliton solution}
The reality requirement of $\f_{c-s}$ was satisfied by taking real $u$ in the case of Type -1A solution.
There is another solution for the reality requirement, which is
by taking $u=w+i K'$, with real $w$ (to be called as the Type-1B solution). 
In the Type-1B case, we obtain following expressions by substituting
$u=w+i K'$ in the corresponding expressions of the Type-1A solution;
\be
\pp \f_{c-s} (z,\zb) ={2 \over k} \sqrt{\b \over V} \dn(\ch,k) +4 k \sqrt{\b \over V} {\sn(w,k) ~\cn(w,k) \over \dn(w,k)}{S \over S^2+1},
\ee
where
\be
S=  {a {\cn(\ch-w,k)  e^M \Theta_s (\ch -w-i K')/ \sn(\ch -w,k)} -b \dn(w,k) e^{-M} \Theta_s(\ch +w+i K') \over
b {\cn(\ch+w,k) e^{-M} \Theta_s(\ch +w+i K')/ \sn(\ch +w,k)}  +a \dn(w,k) e^M \Theta_s (\ch -w-i K')},
\label{S1}
\ee
and
\be
M = -\sqrt{V \b} {\dn (w,k)^4 +k^2 -1 \over k^3 \sn (w,k) ~\cn (w,k)  ~\dn (w,k)} \zb  
+\left( {\Theta_s '(w+i K') \over \Theta_s(w+i K')} +{\cn (w,k) (k^2 \sn (w,k)^2 -2) \over 2 \dn (w,k) \sn (w,k)} \right) \ch.
\label{MN1}
\ee

\subsection{Properties of the solution}
Figure 3(a) plots $\pp \f_{c-s}$ of Type-1B as a function of time $t$ and space $x$.
The parameters used are those of Fig.1, i.e., $V=1, k=0.8, u=0.7, \b=1, a=b=1$.
The distinct difference between the Type-1A and Type-1B solution can be seen
from a time sliced view of the soliton at $t=0$, see Fig. 3(b) and compare it with Fig. 1(b).
The intrinsic velocity $v_0$ of the Type-1B soliton can be obtained from that of the Type-1A soliton
by substituting $u \rightarrow w+i K'$, which gives $v_0 (u=w+i K') =-v_0 (w)$.
Thus the characteristics of the intrinsic velocity $v_0$ of the Type-1B soliton is essentially the same
as those of the Type-1A soliton.
The shift of crests of the
cnoidal wave can be similarly calculated as in the Type-1A solution, which is $4w$.
\begin{figure}
\leftline{\epsfxsize 2.6 truein \epsfbox {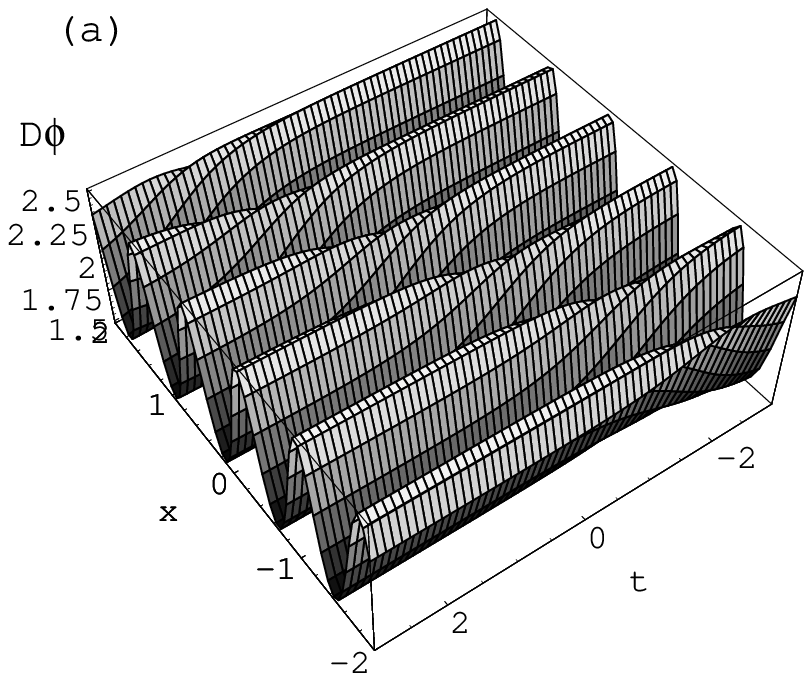}}
\vglue -1.9 in
\rightline{\epsfxsize 2.6 truein \epsfbox {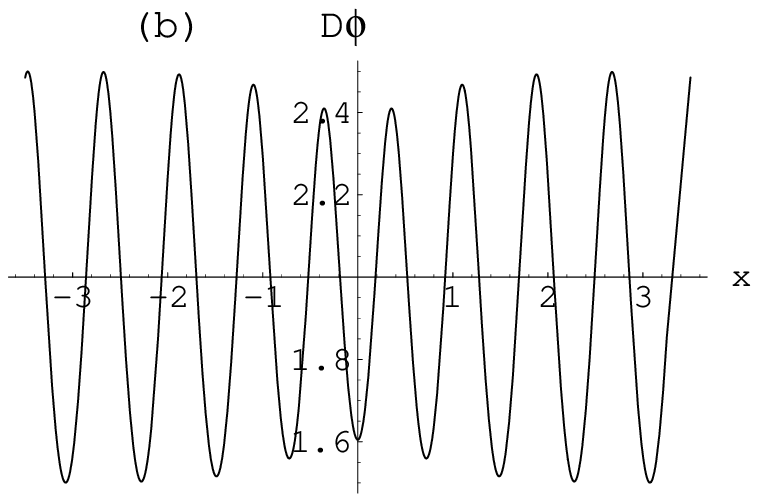}}
\caption{$\pp \f_{c-s}$ showing a sine-Gordon soliton residing on a cnoidal wave background (Type-1B solution);
(a) 3-dimensional plot, (b) time-sliced view at $t=0$. The parameters are $V=1, k=0.8, u=0.7, \b=1, a=b=1$.}
\end{figure}

\section{Type-2 soliton solution on a cn-type background}
\subsection{soliton solution}
The ``soliton + cnoidal wave" solution with the $\f_c^{(2)}$ background can be obtained
by substituting $k \rightarrow 1/k, u \rightarrow ku, \ch \rightarrow k^2 \ch$ in the Type-1A solution.
We call it the Type-2 solution. 
Note that $\dn(\ch,k) \rightarrow \dn(k^2 \ch,1/k)=\cn(k \ch,k)$ such that $\pp \f_c^{(1)}(z, \zb) \rightarrow \pp \f_c^{(2)}(z, \zb)$.
In this case
\be
\pp \f_{c-s} (z,\zb) =2 k \sqrt{\b \over V} \cn(k \ch,k) +4 \sqrt{\b \over V} {\cn(u,k) \over \sn(u,k) ~\dn(u,k)}{S \over S^2+1},
\label{type2}
\ee
where
\be
S=  -{a k \sn(u,k) \cn(k \ch -u,k) e^M \Theta_t(\ch_2 -u) +b \dn(u,k) e^{-M}  \Theta_t(\ch_2 +u) \over
b k \sn(u,k) \cn(k \ch +u,k) e^{-M}  \Theta_t(\ch_2 +u) +a \dn(u,k) e^M \Theta_t(\ch_2 -u)},
\label{S2}
\ee
with
\be
M = \sqrt{V \b} {1-2 k^2 \sn (u,k)^2 +k^2 \sn (u,k)^4 \over \sn (u,k) ~\cn (u,k) ~\dn (u,k)} \zb 
+\left( k{\Theta_t '(u) \over \Theta_t(u)} +{k \cn (u,k) (1-2 k^2 \sn (u,k)^2 ) \over 2 ~\dn (u,k) ~\sn (u,k) } \right) \ch,
\label{MN2}
\ee
and
\be
\Theta_t(u) =\theta_0 \left( {\pi u \over 2 (K-iK')}\right) = 1 + 2 \sum (-)^n q^{n^2} \cos ({n \pi u \over K-iK'}),
\ee
with $q =\exp [-\p K' / (K-iK')]$.

\subsection{Properties of the solution}
Figure 4(a) plots $\pp \f_{c-s}$ of the Type-2 soliton as a function of time $t$ and space $x$.
The parameters used are those of Fig.1, i.e., $V=1, k=0.8, u=0.7, \b=1, a=b=1$.
The distinct difference between the Type-2 and Type-1 solutions can be seen
from a time sliced view of the soliton at $t=0$, see Fig. 4(b) and compare it with Fig. 1(b) and Fig. 2(b).
\begin{figure}
\leftline{\epsfxsize 2.6 truein \epsfbox {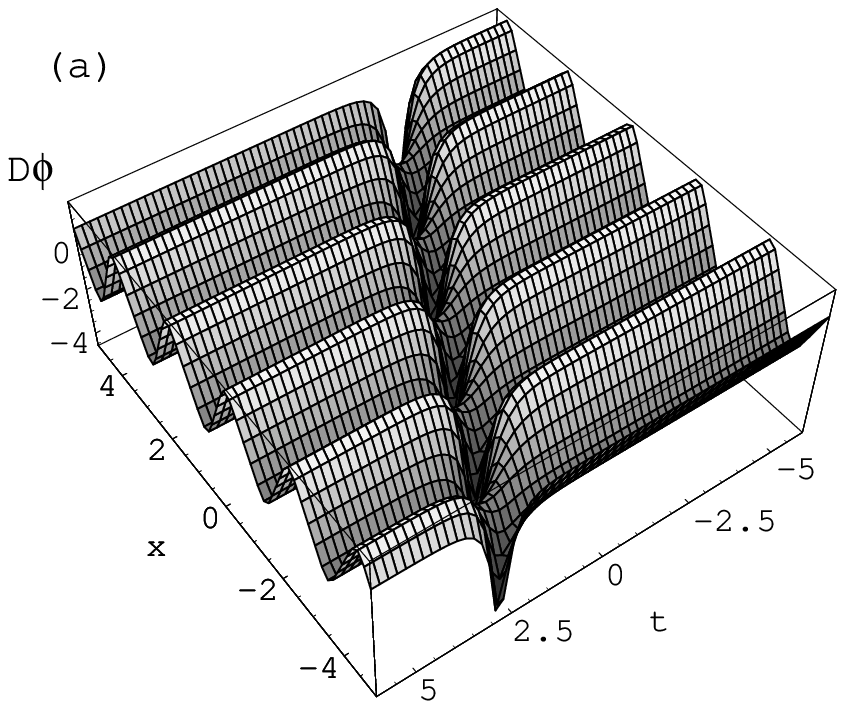}}
\vglue -1.8 in
\rightline{\epsfxsize 2.6 truein \epsfbox {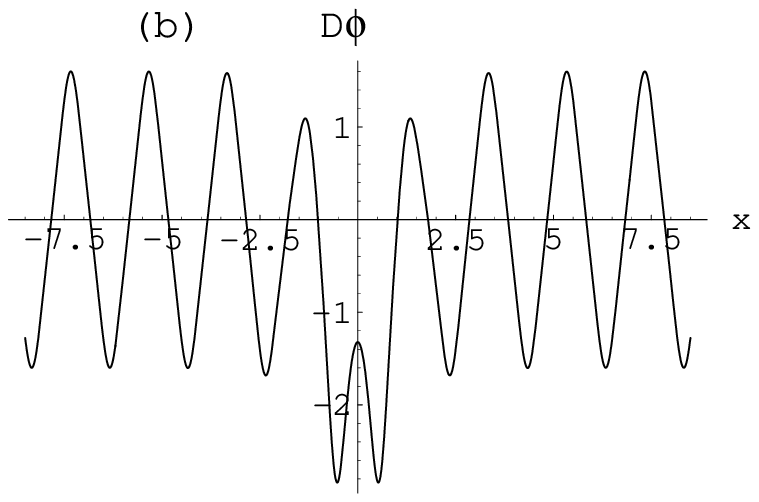}}
\caption{$\pp \f_{c-s}$ showing a sine-Gordon soliton residing on a cnoidal wave background (Type-2 solution);
(a) 3-dimensional plot, (b) time-sliced view at $t=0$. The parameters are $V=1, k=0.8, u=0.7, \b=1, a=b=1$.}
\end{figure}
The intrinsic velocity $v_0$ of the Type-2 soliton can be obtained from that of Type-1A soliton
by substituting $k \rightarrow 1/k, u \rightarrow ku$, which gives 
\be
v_0 ={\dn^2 u ~\sn^2 u +\cn^2 u~(1-2 k^2 \sn^2 u) \over \sn u~\cn u~\dn u (-4 \Theta_w '(u) / \Theta_w(u)
+2 \p u/(K K') )
+\dn^2 u~\sn^2 u-\cn^2 u ~(1-2 k^2 \sn^2 u)},
\ee
where we use the identity
\be
{\Theta_t '(u) \over \Theta_t (u)}=-\p {u \over 2 (K-iK') K'} +{\Theta_w '(u) \over \Theta_w(u)}
\ee
with
\be
\Theta_w (u) =\theta_2 ( -i {\pi u \over 2 K'}) = 2 q^{1/4} \sum q^{n^2+n} \cos ({-i (2n+1) \pi u \over 2 K'}),
\ee
with $q =\exp (-\p K / K')$.
The characteristics of the intrinsic velocity $v_0$ of Type-2 soliton is shown in Fig. 5.
\begin{figure}
\centerline{\epsfxsize 3. truein \epsfbox {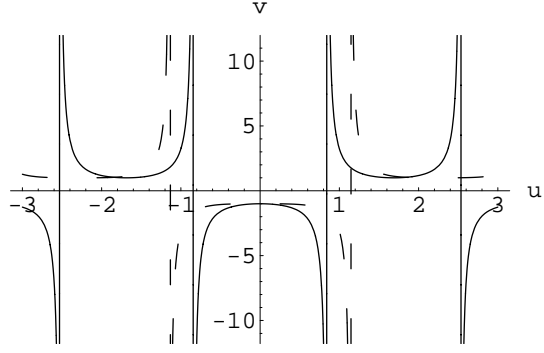}}
\caption{Intrinsic velocity $v_0$ versus $u$ of the Type-2 sine-Gordon soliton for $k=0.5$ (solid line) and $k=0.9$ (dashed line).
.}
\end{figure}
Special values are $v_0 = -1$ at $u=2n K$, n=integer and $v_0=1$ at $u=(2n+1) K$.
At $u=(n+1/2) K$, $v_0 = \pm \infty$. It shows that type-2 solitons move with the velocity $|v| >1$.
We can obtain soliton solutions having $|v| <1$, by
taking the B\"{a}ckrund transformation $ (\f_{c-s}, x, t)  \leftrightarrow  (\f_{c-s}+\pi , t, x)$. 
The shift of crests of the
cnoidal wave can be similarly calculated as in the Type-1A solution, which gives $4 u$ in the Type-2 solution.
The substitution $u = w+i K'$ (used to obtain the Type-1B solution from the Type-1A solution) does not give
new solution in the case of Type-2 solution. In fact, there exists a symmetry $u \leftrightarrow -u+K+i K'$
in the Type-2 solution.

In the $k \rightarrow 0$ limit of the Type-2 solution, $\f_c^{(2)} \rightarrow 0$.
Using the fact that $K(k=0)=\p/2, K'(0)=\infty$, we obtain $S = -\exp(-2 M+\ln b / a)$ where
\be
M =\sqrt{V \b} \left({\sin u \over \cos u} + {\cos u \over \sin u} \right) \zb+{\cos u \over 2 \sin u} k \ch.
\ee
Especially at $u=K/2=\p/4$ with $a=b$ ($\b=V=1$), we obtain
\be
\f_{c-s}=-2 \tan^{-1} \exp(4 t),~ \pp \f_{c-s} = -2 \sech 4t,~\sin 2 \f_{c-s}= 2 \sech 4t ~\tanh 4t.
\label{sol}
\ee
Considering the $x \leftrightarrow t, \b \leftrightarrow -\b$ symmetry of the sine-Gordon theory,
the solution in Eq. (\ref{sol}) becomes the well-known sech-type soliton (without a background) of $\b=-1$ theory.

\section{Discussion}
In this paper, we introduce ``soliton+cnoidal wave" solutions of
the sine-Gordon equation. It was obtained using  the DT and the Sym's solution of the associated linear problem
on a cnoidal wave background. The solutions are expressed in terms of the Jacobi elliptic functions
and can be easily manipulated to obtain interesting physical characteristics. 
In fact, we use the symbolic package MATHEMATICA to check  various formulae of the present paper 
including the solutions itself,
as well as to obtain the figures. We calculate the velocity of a soliton on a cnoidal wave and
the shift of crests of the cnoidal wave. The velocity of the soliton is given by the relativistic
addition of two velocities, the velocity of the cnoidal wave and the intrinsic velocity of the soliton
observed in the moving frame of the cnoidal wave. The intrinsic velocity is determined by the
property of the cnoidal wave ($dn$ or $cn$ wave and the modulus $k$) and the DT parameter $u$.
It would be interesting to make an (numerical) analysis about
the nature of the solitons, especially when it lies on a finite-width cnoidal wave background.

The stability analysis of these solutions is remained for future study. There already appeared that linearized
instabilities of N-phase
solutions can be labeled in terms of spectral data \cite{sin}. It implies that there exist regions
in the parameter space of $u$, $k$ and $L$ (finite length of cnoidal waves), where the solutions are stable.
Explicit relation between the spectral
data and DT parameter $u$ is not understood yet.

\vglue .2in
\centerline{\bf ACKNOWLEDGMENT}
\vglue .2in
This work was supported by Korea Research Foundation Grant (KRF-2003-070-C00011).

\vglue .2in

\end{document}